\documentclass[aps,prl,twocolumn,amsmath,amssymb,nofootinbib,superscriptaddress,floatfix,reprint,longbibliography]{revtex4-2}
\usepackage[T1]{fontenc}
\usepackage{amsmath}
\usepackage{amsfonts}
\usepackage{amsthm}
\usepackage{amssymb}
\usepackage{lipsum}
\usepackage{graphicx}
\usepackage[usenames,dvipsnames]{color}
\usepackage{float}
\usepackage{multirow}
\usepackage{soul}
\usepackage{appendix}
\usepackage[colorlinks=true, urlcolor=blue, linkcolor=blue, citecolor=blue, pdftex]{hyperref}
\usepackage{dcolumn}
\usepackage{bm}
\usepackage{url}
\usepackage{braket}
\usepackage[range-units=single,separate-uncertainty]{siunitx}
\usepackage{verbatim} 
\usepackage{booktabs}
\usepackage[version=4]{mhchem}
\usepackage{adjustbox} 

\usepackage{textcomp}
\usepackage{xcolor}
\usepackage{graphicx}
\usepackage{nicefrac}
\usepackage{amsfonts}
\usepackage[version=4]{mhchem}
\usepackage{amssymb}
\usepackage{amsmath}

\usepackage{subfigure}
\usepackage{multirow}
\usepackage{tabularx}
\usepackage{array}

\usepackage{siunitx}
\usepackage{nicefrac}
\usepackage{braket}
\usepackage{gensymb}

\usepackage{bm}
\usepackage{hyperref}
\usepackage{rotating}

\newcommand{\NIO}{Na$_2$IrO$_3$~}
\newcommand{\CIO}{Cu$_2$IrO$_3$~}
\newcommand{\AIO}{A$_2$IrO$_3$~}
\newcommand{\LIO}{Li$_2$IrO$_3$~}

\begin{document}

\title{Role of Disorder in Governing the Magnetic Properties of Cu$_2$IrO$_3$}


\author{Priyanka Yadav\textsuperscript{1}, Sumit Sarkar\textsuperscript{2}, Vishal Kumar\textsuperscript{3}, Sanjay Singh\textsuperscript{3}, Martin A Karlsen\textsuperscript{4}, Martin Etter\textsuperscript{4}, Sourav Chowdhury\textsuperscript{4}, Subhajit Nandy\textsuperscript{4}, Yogesh Singh}
\affiliation{Department of Physical Sciences, Indian Institute of Science Education and Research (IISER) Mohali, India \\
\textsuperscript{2} Department of Physics and Astronomy, University of Waterloo, Waterloo, Canada\\
\textsuperscript{3}School of Materials Science and Technology, Indian Institute of Technology (BHU), Varanasi-221005, India\\
\textsuperscript{4} Deutsches Elektronen-Synchrotron DESY, Notkestr.85, 22607 Hamburg, Germany \\}

\flushbottom
\begin{abstract}
Cu$_2$IrO$_3$ is a honeycomb iridate which has been studied recently as a candidate Kitaev quantum spin liquid.  Its magnetic ground state however, has been reported to be quantum disordered, spin glassy, or magnetically ordered depending on synthesis details.  We have prepared a Cu$_2$IrO$_3$ sample with large antisite disorder and studied in detail its structure (global and local), charge states, and thermodynamic properties to try to quantify and characterize the disorder and its connection to the magnetic ground state. X-ray diffraction, Extended x-ray absorption fine structure(EXAFS) and X-ray pair distribution function analysis revealed a large site disorder ($\sim 25\%$), while XPS and XANES reveal mixed valence of Cu and Ir following Cu$^{1+}$ + Ir$^{4+}$ $\rightarrow$ Cu$^{2+}$ + Ir$^{3+}$. This combination of site disorder and charge redistribution generates competing antiferromagnetic interactions and magnetic frustration, resulting in dynamically fluctuating AFM clusters near 80 K that freeze below 29 K. These results demonstrate the crucial role of synthesis dependent disorder in determining the magnetic ground state of Cu$_2$IrO$_3$.

\end{abstract}
\maketitle
\section{Introduction}
In recent years, there has been a surge of interest in quasi-two-dimensional (2D) 
honeycomb iridates and ruthenates as promising candidates for realizing the Kitaev quantum spin liquid (QSL) ground state.
The basic ingredients to realize the quantum spin liquid (QSL) state in 
these materials are the presence of strong spin-orbit coupling (SOC), which
gives rise to a pseudo-spin $J_{\text{eff}} = \frac{1}{2}$ state, and
the edge-sharing octahedral network that fosters bond-dependent anisotropic Ising exchange interactions leading to frustration \cite{HKmodel,Kitaev,TREBST,HK}.
The leading candidate materials
such as \NIO, $\alpha$-RuCl$_3$, and $\alpha$-\LIO however, also exhibit additional interactions 
such as isotropic Heisenberg exchange
(often extending beyond nearest neighbors) and off-diagonal exchange \cite{NIO,AIO,RCL} leading to long-range magnetic order
at low temperatures \cite{LIO,RCL,NIO}.
The quest for an ideal Kitaev material  has led to the synthesis of second generation Kitaev materials 
including Cu$_2$IrO$_3$, H$_3$LiIr$_2$O$_6$, and Ag$_3$LiIr$_2$O$_6$. These compounds
are derived from the first generation Kitaev materials $A_2$IrO$_3$, through subtle structural
modifications. For example, H$_3$LiIr$_2$O$_6$ is obtained by replacing the interlayer 
Li$^+$ ions in $\alpha$-Li$_2$IrO$_3$ with H$^+$, while preserving the integrity of the 
honeycomb layer. Whereas in Cu$_2$IrO$_3$, all $A$-site cations in Na$_2$IrO$_3$ are
replaced with Cu, which occupy positions both at the center of the
honeycomb and between the honeycomb layers \cite{PRX,PRL,PRB,CIO_JACS}. 
A key feature of these structures is the increase in interlayer separation where the $c$-axis is elongated
via O-Cu-O, O-H-O, or O-Ag-O dumbbell bonds 
bringing them closer to an ideal configuration with Ir-Ir-Ir
bond angles approaching 120$^\circ$ and Ir-O-Ir angles nearing 90$^\circ$ \cite{CIO_JACS}.
However, in these intercalated compounds, structural disorders play an important role because of the uncontrolled location of intercalated ions whose positions strongly affect the local magnetic interactions \cite{SD} and can obscure the true magnetic ground state \cite{70K,mimic}, sometimes leading to disorder driven phases which mimic quantum spin liquid behavior \cite{mimic,pseudodipole}. 
 An initial study on \CIO revealed local moment behaviour with a spin glass like anomaly at $2.7$~K suggesting weak disorder \cite{CIO_JACS}. Subsequently, studied materials had varying amounts of disorder at Ir and Cu sites in honeycomb layer leading to differences in magnetic behavior at low temperatures (T$ <<\theta_{cw}$).  Whereas a sample with $\sim 5$-$8\%$ disorder showed random singlet behavior in susceptibility measurements at low temperatures and only dynamic fluctuating moments in muon spin relaxation($\mu$SR) \cite{PRL}, a sample with $\sim 20\%$ mixed Cu$^{1+/2+}$ and Ir$^{3+/4+}$ charge states revealed both static and dynamic spins in different volumes of the sample \cite{PRB}. These studies on the one hand point to the possibility of smaller non-Kitaev
interactions leading to the absence of magnetic order and a QSL like state despite its Curie-Weiss temperature and effective magnetic moment being similar 
to those of Na$_2$IrO$_3$ \cite{CIO_JACS, PRL,Spal,PRB}. On the other hand they point to the role of disorder in deciding the magnetic ground state. Recently the magnetic properties of \CIO synthesized
at lower temperature using CuCl-KCl eutectic salts has been reported \cite{70K}. 
Using this method the synthesis temperature was lowered from 350$^\circ$C to 170$^\circ$C.  This  resulted in samples with 5\% Cu/Ir site disorder and magnetic susceptibility revealed a weak ferromagnetic anomaly around $70$~K\@.  These reports emphasize the crucial role of chemical disorder in Kitaev materials.

A recent density functional theory (DFT) study has proposed that the mixed valence could lead to long-range charge order with alternate Ir$^{4+}$ and Ir$^{3+}$ on the honeycomb lattice \CIO \cite{Radu}.
The study also showed that while in the higher symmetry structure
\textit{C}2/\textit{m} crystallographic constraints
suppress the tendency towards charge-order, the lower symmetry \textit{C}2 phase energetically favors charge ordering albeit with a small energy gain $\approx 4$~meV/f.u.  With such a low energy gain,
its stabilization in real materials is likely hindered by factors such as stacking faults, entropy, and geometric frustration. 
Consequently, the system may adopt a disordered or glassy distribution of locally mixed valent
Ir ions, which preserves the average  \textit{C}2/\textit{m} 
symmetry in diffraction measurements \cite{Radu}.
The crystal structure and chemical disorders in the layered materials are highly sensitive to synthesis conditions.
Factors like temperature, synthesis atmosphere, and annealing time can significantly influence phase formation, 
structural integrity, and cation ordering in these systems \cite{Li2IrO3,SR_Radu,Ag3LIO_SD}.

In the present study, we synthesized Cu$_2$IrO$_3$ by employing a prolonged topotactic reaction at 320~$^\circ$C. Structural refinement revealed an increased cation (chemical) disorder of approximately 25\%. The extent and spatial distribution of this disorder were quantified using complementary local structure probes such as extended X-ray absorption fine structure (EXAFS) and X-ray pair distribution function (PDF) analyses. Both techniques provided detailed insight into the local atomic environment, revealing a coexistence of ordered and disordered regions within the honeycomb plane. To quantify the antisite disorder EXAFS and PDF data were modeled using two structural components, one comprising of well-ordered Cu$^{1+}$-Ir$^{4+}$ honeycomb network and other exhibiting Cu-Ir site exchange within the honeycomb lattice.
The degree of disorder was quantified from model fits based on the C2/c crystal structure.
To further probe the charge configuration associated with this disorder, X-ray photoemission spectroscopy (XPS) and X-ray absorption near-edge structure (XANES) measurements were performed. These results confirmed the presence of mixed oxidation states of Cu and Ir, indicating local charge redistribution accompanying the cation disorder.
In the ideal honeycomb structure, 
Ir$^{4+}$ moments experience Kitaev type bond dependent interactions, 
which can drive a quantum spin liquid (QSL) state. However, 
cation site disorder disrupts this framework by introducing Cu-Ir site swapping, 
which leads to an additional source of frustration.
Specifically, when Cu and Ir swap positions, local triangular motifs can
emerge within the lattice leading to additional
exchange pathways of the type Cu$^{2+}$-O-Ir$^{4+}$ in addition to the usual Ir$^{4+}$-O-Ir$^{4+}$.
This potentially introduces competing magnetic interactions that contribute to spin frustration.
Indeed, magnetic measurements revealed
the impact of this disorder on the macroscopic magnetic behavior. 
DC susceptibility measurements demonstrated spin-glass-like
freezing below $29$~K, while a frequency dependent peak at $\approx$ 80 K in the AC susceptibility suggests slow spin dynamics consistent with cluster glass formation already at high temperatures.

\section{Experimental Methodology}
\CIO has been synthesized from its parent material \NIO by using a topotactic reaction \cite{CIO_JACS}. The 
ion exchange reaction \NIO$+~2$CuCl $\rightarrow$ Cu$_2$IrO$_3$ + 2NaCl replaces the Na$^+$ with Cu$^+$ ion. 
\NIO and CuCl were mixed in a molar ratio of 1:2.2 using a mortar and pestle and then sealed in a quartz tube in an argon atmosphere. The sealed sample 
 was annealed in a muffle furnace for 48 hours at 320$\degree$C temperature and then slowly cooled to room temperature at a rate of
 1$\degree$C per minute.
The \CIO compound is separated from NaCl by washing it in ammonium hydroxide (NH$_4$OH, Alfa
Aesar, 28$\%$) and DI water. 
The phase purity and stoichiometry of \CIO was characterized by synchrotron x-ray diffraction (XRD) with \(\lambda = 0.20737 \text{\AA}\) 
at Powder Diffraction and Total Scattering Beamline P02.1 at the PETRA III synchrotron at DESY, Hamburg, Germany.
The relative stoichiometric ratio Cu:Ir has also been confirmed using 
energy dispersive x-ray spectroscopy (EDX) mapping with a scanning electron microscope. 
Local coordination environment was probed using EXAFS
at Ir L$_3$ edge and Cu K edge.
The EXAFS spectra were recorded in flourescence mode 
at Advanced X-ray Absorption Spectroscopy Beamline P64 at the PETRA III synchrotron at DESY, Hamburg, Germany\cite{exafsexp}. 
Reference absorption edge spectra of materials 
IrO$_2$ and CuO were used for energy calibration 
of incident X-ray in XANES and EXAFS measurement. In order to get better statistics, the EXAFS scans 
were collected four times and averaged.
Normalized XANES spectra and EXAFS
oscillations in \( K \)-space, \( \chi(K) \),
were extracted following standard procedures using
the ATHENA program\cite{Ravel}. The EXAFS signal, within a selected \( K \)-range, 
was Fourier-transformed to obtain the corresponding \( R \)-space
spectra, \( \chi(R) \). The extracted EXAFS spectra 
were then analyzed using the ARTEMIS software, which employs 
ATOMS and FEFF6 to generate theoretical spectra by
summing contributions from all possible scattering 
paths in a given crystallographic structure. 
Each contribution, such as from the \( i \)th coordination shell,
was modeled using the standard EXAFS equation with 
refinable structural parameters, including the
coordination number (\( N_i \)), average
bond length (\( R_i \)), and the mean-square 
relative displacement (MSRD) factor (\( \sigma^2_i \)). 
These structural parameters were iteratively refined to 
achieve the best fit between experimental and theoretical spectra.
All spectroscopic measurements were performed at 300 K.

The chemical valence state of the elements present in the
\CIO were investigated by X-ray photoelectron spectroscopy
(XPS) experiments using Al K$\alpha$ (h$\nu$ = 1486.7 eV) laboratory source
and hemispherical energy analyzer (Omicron, EA-125,
Germany) at the angle integrated photoemission spectroscopy
(AIPES) beamline (Indus-1, BL 2, RRCAT, Indore, India).
During XPS measurements, experimental chamber
vacuum was of the order of 10$^{-10}$ Torr. The charging effect
corrections in XPS were done by measuring the C 1s
core level spectra. XPS spectra were deconvoluted by fitting
with the combined Lorentzian-Gaussian function and Shirley
background using XPSPEAK 4.1 program.Soft X-ray absorption spectroscopy has been performed at the MAX-P04 end-station, P04 beamline, Petra III, DESY, Hamburg, Germany. The data was measured with total electron yield (TEY) mode by recording the sample drain current.
Room temperature X-ray total scattering measurements
were collected at P02.1 (DESY, Hamburg, Germany) with a wavelength of 0.0.20737\text{\AA} and 
a maximum momentum transfer $Q_{\text{max}} = 25~\text{\AA}^{-1}$. The pair distribution function (PDF) was
obtained by Fourier transforming the total scattering intensity using the PDFgetX3 program\cite{pdfgetx3}.
Structural refinements were performed using PDFgui\cite{pdfgui}.
DC and AC susceptibility measurements were performed using the Quantum design 9T Evercool (II) Physical Property
measurement system (PPMS). 
\section{Results and Discussions}
\subsection{X-ray diffraction and elemental analysis}
\begin{figure}[h]                                                               
\centering                                 
\includegraphics[width=0.5\textwidth]{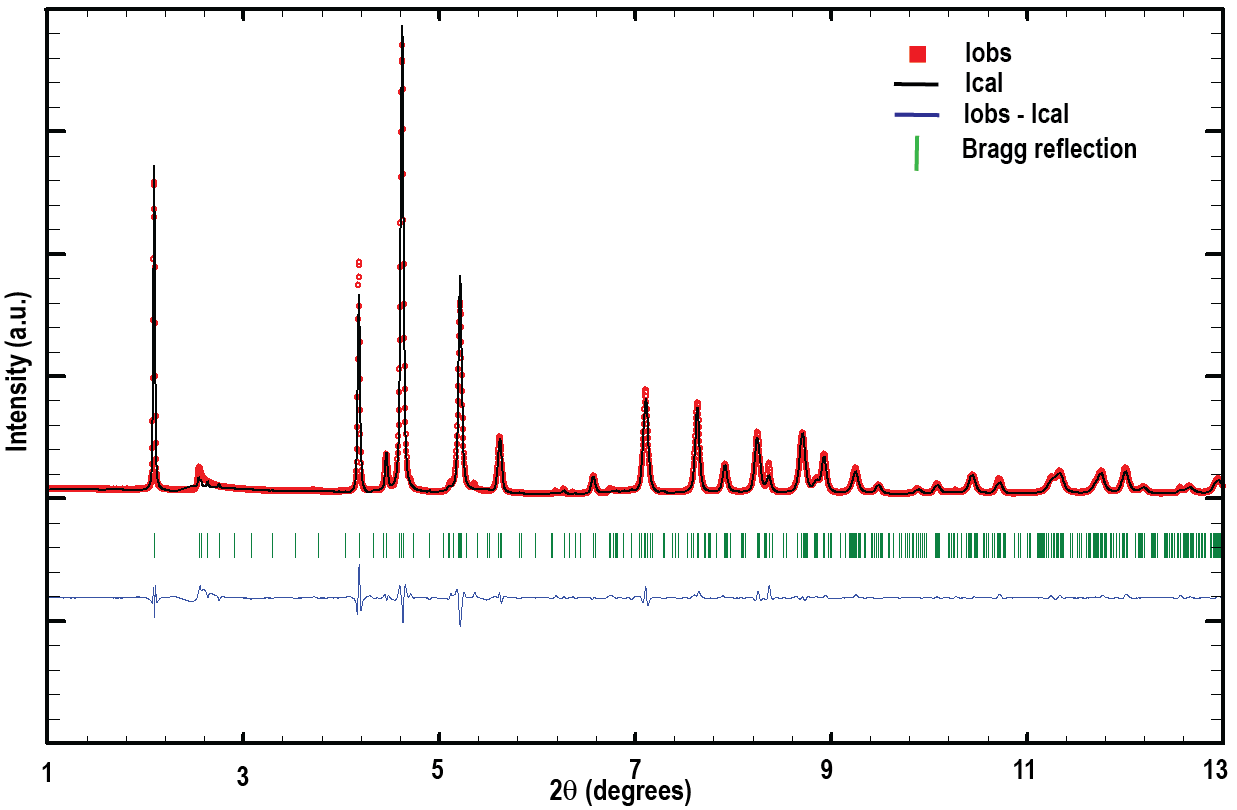}
	\caption{Rietveld refinement of polycrystalline \CIO using X-ray data collected at $\lambda = 0.207371\,\text{\AA}$, confirming that the synthesized powder sample contains no impurity phases.}
\label{XRD}
\end{figure}

A Rietveld refinement of powder XRD (PXRD) shown in Figure.~\ref{XRD} 
confirmed that the synthesized \CIO is single phase. The space groups C2/c, P21/c, C2 and C2/m have been used in previous literature as structural models for \CIO \cite{structures,Radu}. We attempted a Rietveld refinement of our PXRD data using the above space groups and found that the C2/c space group (No. 15) gives the best fit.
The characteristic asymmetric broadening
of peaks between 2.43 to 3.43 $\text{\AA}$ similar to \AIO (A: Li, Na) \cite{NIO,LIO} was observed and indicates the presence of stacking faults.
The refinement gave the lattice parameters $a = 5.3964$~\AA, $b = 9.3373$~\AA, $c = 11.5204 $~\AA, and $\beta = 99.2522$~\degree. 
The refined fractional atomic coordinates, 
occupancies, isotropic thermal factors, and 
the reliability parameter Bragg R-factor 
obtained is given in Table \ref{refinement}. We note that the best refinement gives $\approx 25\%$ antisite disorder between Ir and Cu sites in honeycomb layer. 

\begin{table}[H]
    \centering
    \caption{Refined atomic positions, and occupancies signifying best fit with C2/c space group}
\label{refinement}
	\begin{tabular}{lcccccc}
        \toprule
        Atom & x & y & z & Occupancy &  Bragg R-factor \\
        \midrule
        Ir1  & 0.2627(4) & 0.0809(2) & 0 & 0.75  &  \\
        Cu1  & 0.2627(4) & 0.0809(2) & 0 & 0.25  & 3.311 \\
        Ir2  & 0.75 & 0.25 & 0 & 0.5  &  \\
        Cu2  & 0.75 & 0.25 & 0 & 0.5  &  \\
        Cu3  & 0 & 0.7334(8) & 0.25 & 1.0  &  \\
        Cu4  & 0 & 0.4455(6) & 0.25 & 1.0  &  \\
        Cu5  & 0 & 0.0803(2) & 0.25 & 1.0  &  \\
        O1   & 0.9491(9) & 0.75 & 0.0938(6) & 1.0  &  \\
        O2   & 0.9492(5) & 0.4400(2) & 0.0808(6) & 1.0  &  \\
        O3   & 0.9313(5) & 0.0841(2) & 0.0880(11) & 1.0  &  \\
        \bottomrule
    \end{tabular}
\end{table}

An elemental analysis of the polycrystal \CIO with energy dispersive x-ray analysis using a scanning electron microscopy 
gave the Cu: Ir ratio 1.85:1.  It is worth noting that no Na was detected in the elemental mapping, suggesting that the parent \NIO phase was fully converted into \CIO.

\section{Core level electronic structure}
\subsection{X-ray Photoemission and absorption spectroscopy}
\begin{figure}[h] 
\centering
\includegraphics[width=0.45\textwidth]{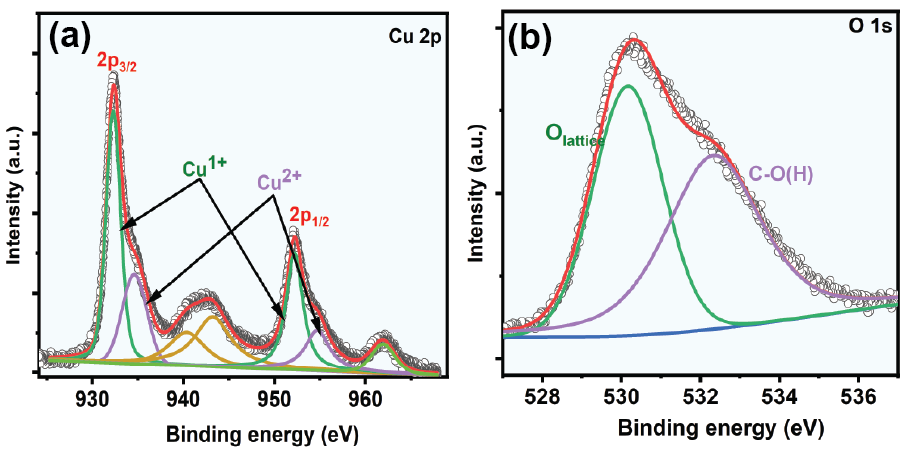}
\caption{The deconvoluted Cu 2p and O 1s core-level X-ray photoelectron spectra (XPS) of \CIO confirming mixed valence state of Cu.}
\label{XPS}   
\end{figure}

We have investigated the elemental charge states in Cu$_2$IrO$_3$ using X-ray photoemission and soft X-ray absorption measurements. The Cu 2p
core-level X-ray photoemission spectra was recorded. Figure \ref{XPS}(a) shows the fitted 
Cu 2p core splitted as 2p$_{1/2}$ and 2p$_{3/2}$ due to the spin-orbit splitting. The observed asymmetry and broadening of the spectral features indicate the presence of
multiple valence states of Cu. The Cu 2p spectrum was deconvoluted into 
components centered at 932.2, 934.6, 940.3, 943.2, 952.1, 954.7, and 961.9~eV.
The peaks at 932.2 and 952.1~eV correspond to the Cu$^{1+}$ state, while those at 934.6 and 954.7~eV 
are attributed to Cu$^{2+}$. The prominent satellite features at 940.3 and 943.2eV (Cu 2p$_{3/2}$ region) 
and 961.9eV (Cu 2p$_{1/2}$ region) are characteristic signatures of Cu$^{2+}$ \cite{Hemant,XPS,XPS2,XPS3}. 

The O 1s core-level spectrum was also recorded. Figure \ref{XPS}(b) shows
a main peak at 530.2~eV, corresponding to lattice oxygen in \CIO, 
along with a higher binding energy feature at 532.2~eV most likely associated with 
chemisorbed oxygen species such as C–O(H)\cite{Hemant}.

\begin{figure}[h]                        
\centering                                   
\includegraphics[width=0.49\textwidth]{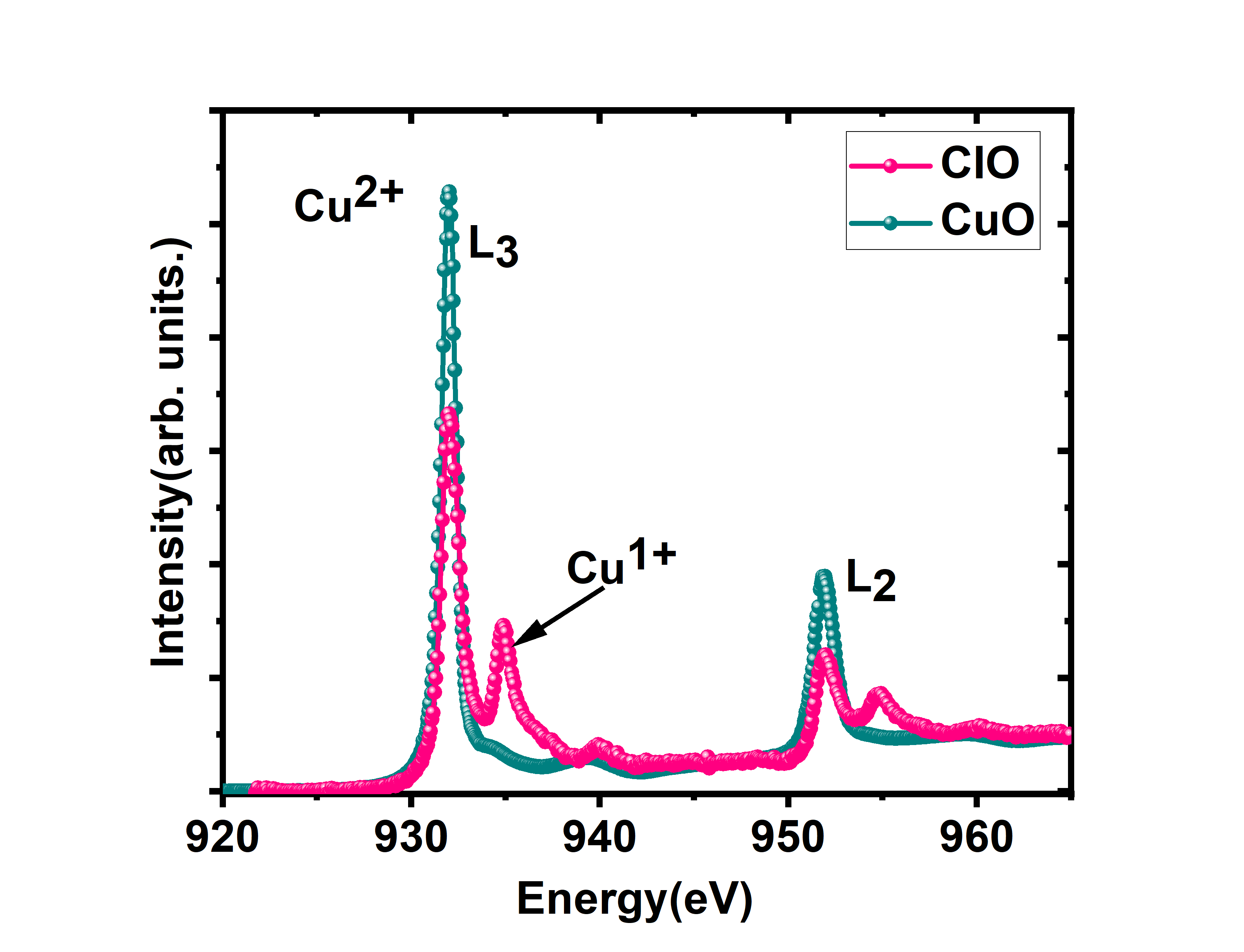} 
\caption{X-ray absorption spectra at the Cu L-edge for \CIO and reference CuO. The comparison clearly demonstrates the presence of both Cu$^{1+}$ and Cu$^{2+}$ oxidation states in \CIO. Distinct spectral features corresponding to the unoccupied 3d states of Cu$^{2+}$, along with those characteristic of Cu$^{1+}$, are observed and are marked with arrows in the figure.}
    \label{XANESLedgeCu}
\end{figure}

To further confirm this mixed valence nature of Cu observed using photoemission spectra, we measured the XANES spectra across Cu L$_3$ and L$_2$ edges, which correspond to the 2p$_{3/2} \rightarrow$ 3d and 2p$_{1/2} \rightarrow$ 3d electronic transitions, respectively. As shown in Fig.~\ref{XANESLedgeCu}, the L$_3$ edge spectra of \CIO shows two prominent peaks at 932~eV and 934~eV. The peak at 932~eV closely matches the peak for CuO reference, indicating the presence of Cu$^{2+}$, while the peak at 934~eV is attributed to Cu$^{1+}$ \cite{PRB,Hemant}. Similarly, at the L$_2$ edge, features are observed at 951.9~eV and 954.9~eV, corresponding to Cu$^{2+}$ and Cu$^{1+}$, respectively. The coexistence of these features provides clear evidence of a mixed-valent state of copper in \CIO.

\section{Local structure}
\subsection{Extended x-ray absorption spectroscopy}
\begin{figure}[h]
 \centering
\includegraphics[width=0.5\textwidth]{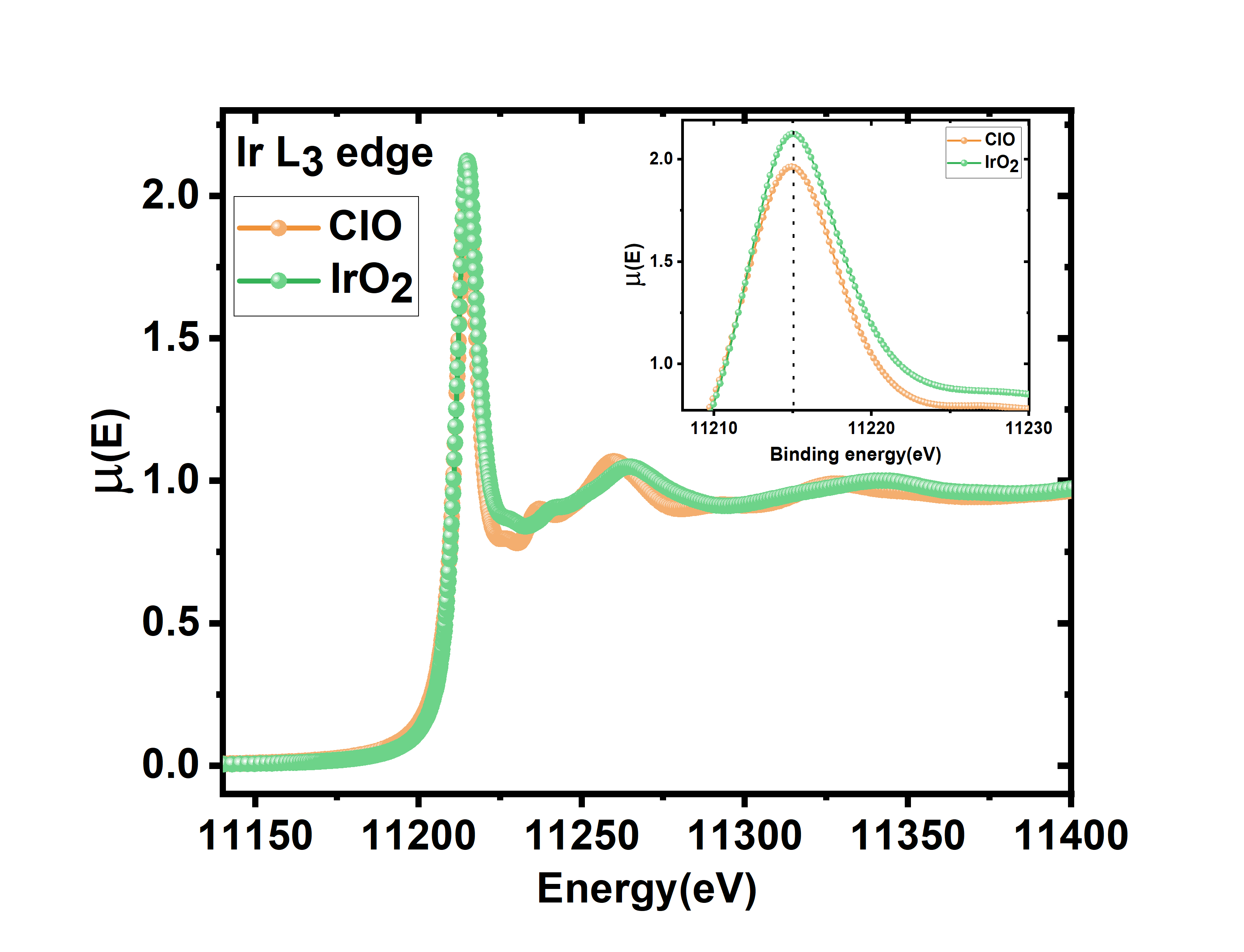}
\caption{Normalized XANES spectra at the Ir L$_3$-edge, with the inset showing a magnified view
	of the white-line feature. Vertical dashed lines serve as visual guides.}
 \label{Irexafs}
\end{figure}

EXAFS measurements at the Ir L$_3$ and Cu K edges were carried out to probe the local coordination environment, yielding quantitative information on interatomic distances, coordination numbers, and short-range structural disorder. These measurements also enabled determination of the oxidation states of both Ir and Cu.
EXAFS is best suited to probe the local effect around scattering center and is sensitive to 
an order below \AA ~scale \cite{EXAFS1,supriyoda}. X-ray absorption 
coefficient as a function of incident photon energy for Ir L$_3$ edge is shown in Fig.~\ref{Irexafs}. 
The obtained spectra can be divided into two regions: X-ray Absorption Near-Edge Spectroscopy (XANES) 
and Extended X-ray Absorption Fine Structure (EXAFS).
The XANES region corresponds to the excitation of electrons from the
2p core level to the unoccupied 5d orbital.

\begin{figure}[h]
\centering
\includegraphics[width=0.45\textwidth]{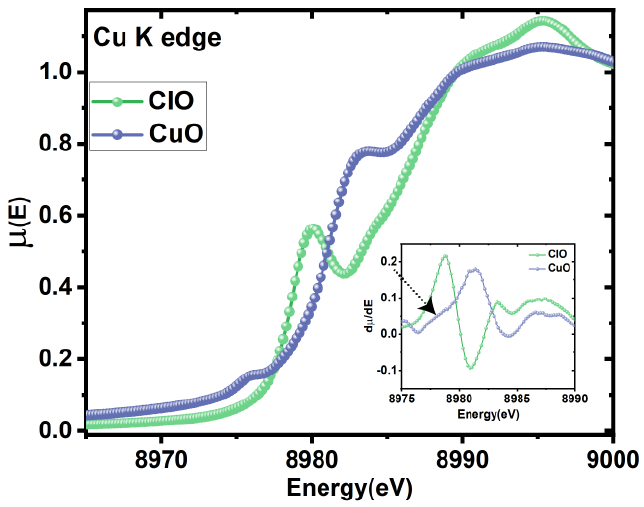}
\caption{Normalized XANES spectra at the Cu K-edge, with the inset 
	highlighting the inflection points of the Cu K-edge spectra 
	used to determine the Cu valence state. The analysis indicates 
	that Cu in \CIO\ is predominantly in the +1 oxidation state,
	with a minor contribution from Cu$^{2+}$, as confirmed by comparison with reference CuO (marked arrow).}
\label{Cuexafs}
\end{figure}

The spectrum of a pure IrO$_2$ sample in which iridium
exists in the 4+ charge state is used as a reference and was
simultaneously recorded with Cu$_2$IrO$_3$. The recorded spectrum 
revealed that the whiteline feature for \CIO is negative shifted compared to IrO$_2$ L$_3$ edge as shown in the inset of Fig.~\ref{Irexafs}.
Such a shift indicates the presence of a mixed Ir$^{3+}$/$^{4+}$ charge state \cite{EXAFSIr3+,Cheng_2016,PRM,PY}.
Cu K edge is also recorded to probe the multivalency of Cu in Cu$_2$IrO$_3$.
CuO reference with oxidation state 2+ is measured along with Cu$_2$IrO$_3$. Figure \ref{Cuexafs} shows the XANES spectra 
of \CIO and CuO. Inset shows derivative of absorption coefficient,
marked arrow denotes Cu$^{2+}$ charge states in \CIO along with Cu$^{1+}$ charge states \cite{Cuexafs,Cuexafs2}.
To explicitly investigate the impact of site disorder, a quantitative 
analysis of $|\chi(R)|$ was performed by modeling the EXAFS data 
based on the \CIO crystal structure (\textit{C2/c}). The fits were 
constrained to an $R$ range of $1 \text{\AA} < R < 3.5 \text{\AA}$ and
a $k$ range of $3.0 \text{\AA}^{-1}< k < 14.5 \text{\AA}^{-1}$. 
Within this region, the EXAFS spectra primarily arise from 
photoelectron scattering by the nearest-neighbor octahedral
O atoms and the second-nearest-neighbor Cu/Ir atoms, 
which are linked to the Ir core absorber through
intermediate O atoms. Based on statistical significance,
the most relevant single and multiple scattering paths
were incorporated into the theoretical EXAFS model. Along with the Ir-O and Ir-Ir/Cu single-scattering contributions, the Ir-O-O multiple-scattering pathway also contributes significantly.

To define the atomic site disorder (ASD) parameter in the fitting model, 
two types of structural configurations are considered,
both centered around Ir atoms. In the first case, all Ir atoms occupy positions 
within a hexagonal plane, with Cu atoms located at the center of the hexagonal plaquette,
representing a perfectly ordered state. In the second case, Cu and Ir atoms in the honeycomb plane  
fully exchange sites, and this structure is treated as the structural limit corresponding to 100\% antisite disorder. 
The theoretical EXAFS spectrum is simulated as a convolution 
of contributions from these two structural configurations \cite{supriyoda,DD}.  

To determine the fractional contribution from each configuration,
the coordination numbers of Ir-Ir$_o$,Ir-Cu$_o$, Ir-Ir$_d$ and Ir-Cu$_d$ bonds in both ordered and disordered 
phase are refined, Ir$_o$, Cu$_o$, Ir$_d$, and Cu$_d$ denote Ir in ordered configuration, Cu in ordered configuration, 
Ir in disordered configuration and Cu in disordered configuration, respectively.

The \(\zeta_{\text{ASD}}\) parameter quantifies the probability of disordered bond configurations and is expressed as:  

\begin{equation}  
\zeta_{\text{ASD}} = 1 - q,  
\end{equation}  

where 

\begin{equation}
	q = \frac{N_{\text{Ir-Ir$_o$/Cu$_o$}}}{N_B}, \quad N_B = N_{\text{Ir-Ir$_o$}} + N_{\text{Ir-Cu$_o$}} +N_{\text{Ir-Ir$_d$}} + N_{\text{Ir-Cu$_d$}}.
\end{equation}
is probability of ordered bond configuration.
Here, $N_{\text{Ir-Ir$_o$}}$, $N_{\text{Ir-Cu$_o$}}$, $N_{\text{Ir-Ir$_d$}}$ and $N_{\text{Ir-Cu$_d$}}$
represent the coordination numbers 
of ordered (Ir-Ir$_o$/Cu$_o$) and disordered (Ir-Ir$_d$/Cu$_d$) bond configurations,
respectively, and $N_B = 6$ corresponds to total  coordination around the B-site cation, which can be occupied by Ir or Cu under antisite disorder. For \CIO samples, the total coordination numbers were fixed to crystallographic values.

The amplitude reduction factor ($S_0^2$) was kept constant at 0.9 \cite{PY1},
a value obtained from the refinement of the Ir L$_3$-edge EXAFS spectra of IrO$_2$.
The energy shift ($\Delta E_0$) was also kept consistent across all
coordination shells. During refinement, the average coordination distances
and mean-square relative displacement (MSRD) factors were optimized. 
EXAFS is very well modelled using a single Debye-Waller
factor and R parameters for each of these Ir-Ir$_o$/Cu$_o$ and Ir-Ir$_d$/Cu$_d$
scattering paths.

\begin{figure}[h]
\centering
\includegraphics[width=0.5\textwidth]{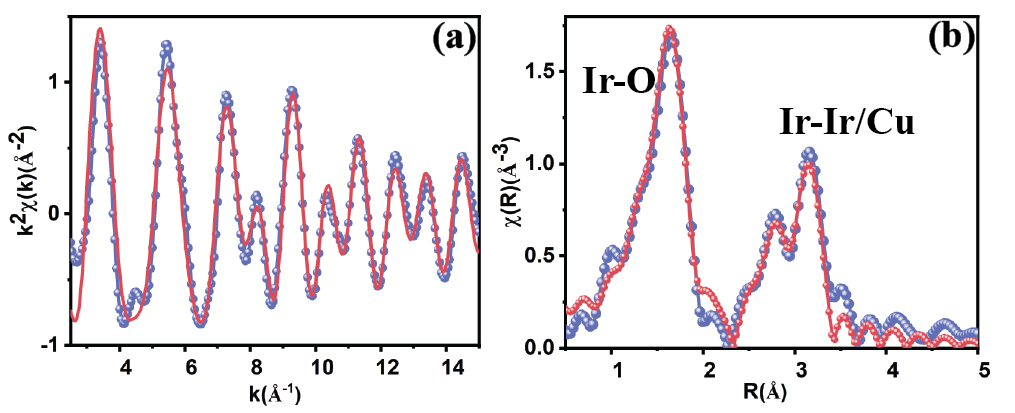} 
\caption{Ir L$_3$-edge extended X-ray absorption fine structure (EXAFS) analysis, where the observed data are shown as blue circles and the best-fit results as solid pink lines in both panels: (a) $k^2$-weighted spectra [$k^2 \chi(k)$] and (b) corresponding Fourier transform modulus [$|\chi(R)|$] for the \CIO bulk sample. Contributions from different coordination shells are identified and distinguished across multiple regions of the spectra.}
\label{EXAFS}
\end{figure}

\begin{figure}[h]  
\centering                                             
\includegraphics[width=0.4\textwidth]{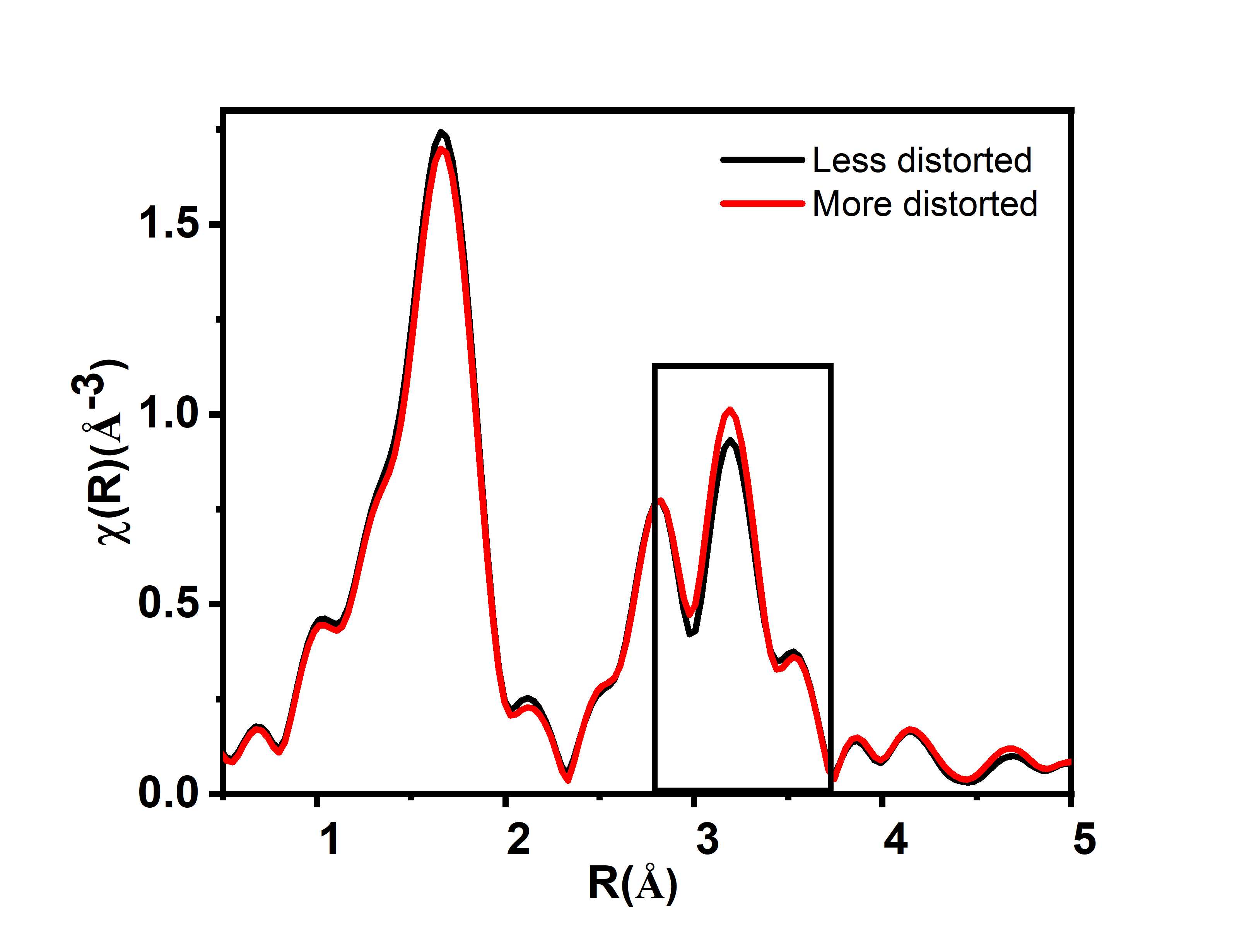} 
\caption{Fourier transform modulus [$|\chi(R)|$] for a more disordered sample (studied here) 
and a less disordered sample (studied earlier). The variation in second-shell intensity 
confirms differences in the fractions of Cu/Ir site mixing.}
\label{comp}
\end{figure}

The quality of fit was assessed using the R factor, defined as:

\begin{equation}
R = \frac{\sum_{i} \left[ \text{Re} (\chi_d(R_i) - \chi_m(R_i))^2 + \text{Im} (\chi_d(R_i) - \chi_m(R_i))^2 \right]}{\sum_{i} \left[ \text{Re} (\chi_d(R_i))^2 + \text{Im} (\chi_d(R_i))^2 \right]},
\end{equation}

where $\chi_d(R)$ and $\chi_m(R)$ are
the experimental and theoretical EXAFS signals, 
respectively \cite{kelly,supriyoda}. The best-fit results are superimposed 
on $k^2\chi(k)$ and $|\chi(R)|$ in Fig.~\ref{EXAFS}(a) and \ref{EXAFS} (b). 
The obtained model patterns show good agreement with the experimental data, 
as confirmed by the goodness of fit indicator yielding a value of approximately 0.01.
The model fitting for the
local structure gave bond lengths as Ir-O = 1.938 \text{\AA}, Ir-Ir$_o$ = 3.177\text{\AA}, Ir-Cu$_o$ = 3.130\text{\AA},
Ir-Ir$_d$ = 3.067\text{\AA}, Ir-Cu$_d$ = 3.067\text{\AA}
while the estimated ASD fractions for the sample is 25\%.
We have also recorded the EXAFS of a  sample with slightly less ASD 18\%  prepared under 
similar synthesis condition as reported by Abramchuk et.al. \cite{CIO_JACS}. 
Figure \ref{comp} compares the $|\chi(R)|$ of the samples with 25\% and 18\% ASD.
The first peaks show no significant difference between 
the two patterns, as they primarily correspond to the
first nearest-neighbor Ir-O bonds. However, a noticeable change
appears in the second peak around 3.2~\AA, which has contributions from Ir-Ir and Ir-Cu bonds. 
The intensity variation between the samples with different amounts of disorder  confirms a different neighborhood for the central absorber Ir and hence points to the different fractions of Cu/Ir site mixing in the two samples \cite{exafsint}.
\\
\subsection{Pair Distribution Function}
\begin{figure}[h] 
\centering
\includegraphics[width=0.50\textwidth]{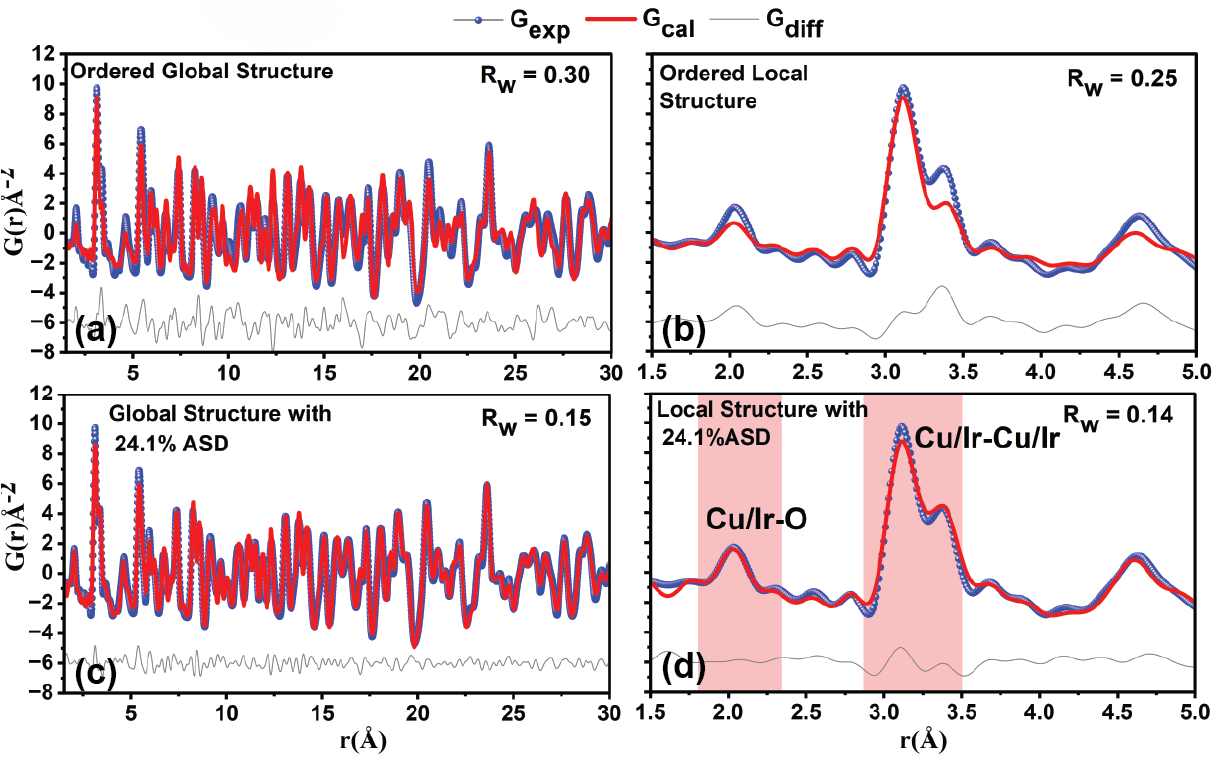}
\caption{Room-temperature X-ray PDF of \CIO tried to fit with ordered structure and ASD. Panels (a,b) show fits using the ordered C2/c model over global (1.5–30 Å) and local (1.5–5 Å) ranges, while (c,d) include 24.1\% ASD. $G_{\text{exp}}$ denotes the experimental PDF, $G_{\text{cal}}$ denotes the calculated PDF from the structural model, and $G_{\text{diff}}$ is the difference curve ($G_{\text{exp}} - $ $G_{\text{cal}}$). $R_{w}$ represents the weighted residual value that quantifies the quality of the fit.
 Incorporating ASD markedly improves the fit (lower $R_w$) and better reproduces the local Cu/Ir–O and Cu/Ir–Cu/Ir structures.}
\label{CIO-HDPDF}   
\end{figure}

To further investigate the local structure and validate the degree of ASD obtained from EXAFS analysis, we performed X-ray pair distribution function (PDF) measurements. This technique is uniquely capable of probing both local and average structural environments. Unlike conventional X-ray diffraction, which yields only the long-range averaged structure from Bragg reflections, PDF analysis incorporates both Bragg and diffuse scattering contributions, enabling the detection of subtle local distortions and antisite cation disorder over the short to intermediate length scales \cite{PDFbook,pdf2,pdf3,pdf4}. In this work, the PDF data were analyzed over a range extending from 1.5 \text{\AA} to 30 \text{\AA}.
 To quantify the ASD the refinement is first performed with the ideal ordered structural model and then the same data is refined with disordered model (structure obtained from PXRD).  Initial long-range refinements (up to 30 \text{\AA} ) were performed to extract the average structure parameters. These parameters were then fixed during the site-occupancy refinements. The Cu and Ir atoms sharing the same crystallographic site were constrained such that their occupancies summed to unity. For the PDF refinements of the disordered structural model, the Cu and Ir occupancies on the mixed 8f site were allowed to be refined simultaneously but with site-multiplicity constraints: if the Cu occupancy on the 8f site is defined as occ(Cu), the corresponding Ir occupancy on the same site was fixed to 1-occ(Cu). This constraint was applied to all symmetry-equivalent positions of the mixed site. Under this restriction, a fully ordered structure corresponds to zero Cu/Ir ASD. Figure~\ref{CIO-HDPDF}(a) shows the refinement using the ordered structural model, which yields a weighted residual factor $R_w = 0.30$. Upon introducing Cu/Ir ASD, $R_w$
 is significantly reduced to 0.15, indicating a substantial improvement in the quality of the fit depicted in Fig.~\ref{CIO-HDPDF}(c). 
Similarly, Fig.~\ref{CIO-HDPDF}(b) presents the PDF refinement of the local structure assuming zero ASD, yielding a weighted residual factor 
$R_w$=0.25. After introducing Cu/Ir ASD, the local-structure PDF refinement shown in Figure \ref{CIO-HDPDF}(d) results in a further reduction of 
$R_w$ to 0.14, indicating a noticeable improvement in the quality of the fit. Since Ir (Z = 77) and Cu (Z = 29) differ significantly in atomic number, their X-ray atomic form factors (f(Q))  differ strongly, providing sufficient scattering contrast to allow the Cu/Ir site occupancies to be distinguished in the PDF.  A pronounced reduction in the weighted residual factor ($R_w$) upon allowing Cu/Ir site mixing to approximately 24.1\% (Figure \ref{CIO-HDPDF}), relative to the fully ordered model, supports the amount of disorder estimated previously from PXRD and EXAFS modeling. 
The peak positions in the real-space G(r) represents the probability of finding an atomic pair separated by a distance r\cite{PDFbook}. 
The shape of a PDF peak provides insight into the distribution of interatomic distances for a given atomic pair, whereas its integrated intensity is related to the coordination number\cite{PDFbook}.
The pink shaded region in Figure \ref{CIO-HDPDF}(d) demonstrates the distribution of coordination pairs in \CIO.

\section{Magnetic properties}
\subsection{DC Magnetisation}
\begin{figure}[h]
    \centering
    \includegraphics[width=0.55\textwidth]{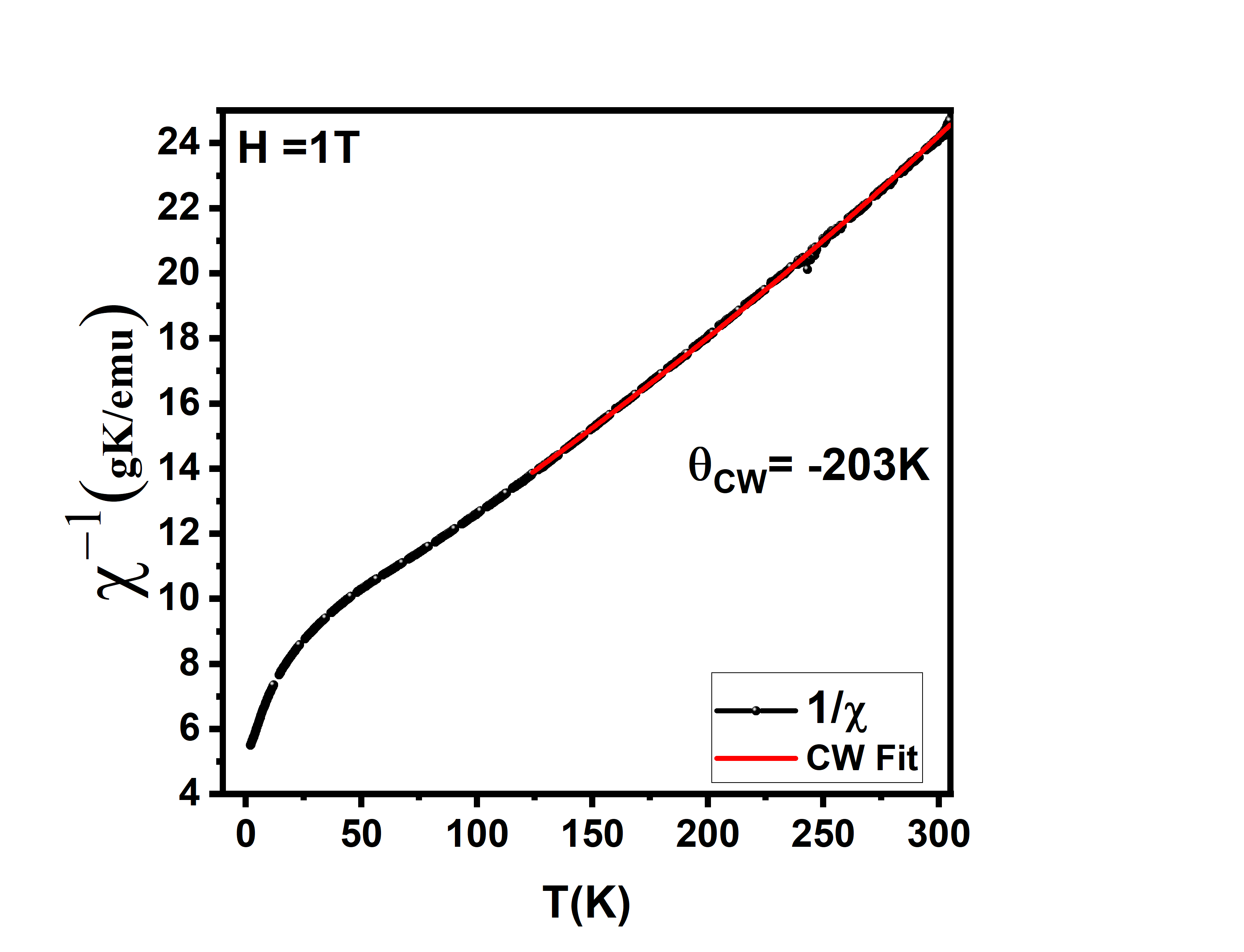}
        \caption{The temperature dependence of dc magnetic
	susceptibility of Cu$_2$IrO$_3$, plotted as $\chi^{-1}$(T) in the temperature
range 2-300 K, measured in a field of 1 T. The solid line represents
the fit to Curie-Weiss law in the range 125 K to 300 K.}
    \label{inversesusceptibility}
\end{figure}

Figure~\ref{inversesusceptibility} shows the inverse susceptibility \(1/\chi\) measured at $H = 1$~T\@. At high temperatures ($T > 125$~K) the data can be fit by the Curie-Weiss (CW) law given by  

\begin{equation}
\chi = \chi_0 + \frac{C}{T - \theta_{\text{CW}}}
\end{equation}

where the temperature independent term \(\chi_0\) accounts for Van Vleck paramagnetism and diamagnetism of core shells,
while \(C\) and \(\theta_{\text{CW}}\) represent the Curie constant and 
Curie-Weiss temperature, respectively. The fit shown as the curve through the data in Figure.~\ref{inversesusceptibility} yields \(C = 28.8\)~emu/g, \(\chi_0 = -0.016\)~emu/g~K 
and \(\theta_{\text{CW}} = -203\)~K\@. The negative \(\theta_{\text{CW}}\) indicates
the dominance of antiferromagnetic (AFM) interactions in the system.
The Curie constant gives an effective magnetic moment \(\mu_{\text{eff}}\) \(\sim 2.38\ \mu_B\), assuming a $g$-factor $g = 2$.  This moment is much larger than the value $1.73~\mu_B$ expected for Ir$^{4+}$ ion with \( J = \frac{1}{2} \).
This indicates a contribution from Cu$^{2+}$ with $S = 1/2$ although we should be careful in trying to quantify the amount of Cu$^{2+}$ from this enhanced moment value because in 5d materials strong spin orbit coupling may lead to a larger $g$-factor which could also lead to an enhanced effective moment compared to the spin-only value. \cite{PY1,HK,mixj1232}.

\begin{figure}[h]
    \centering
    \includegraphics[width=0.49\textwidth]{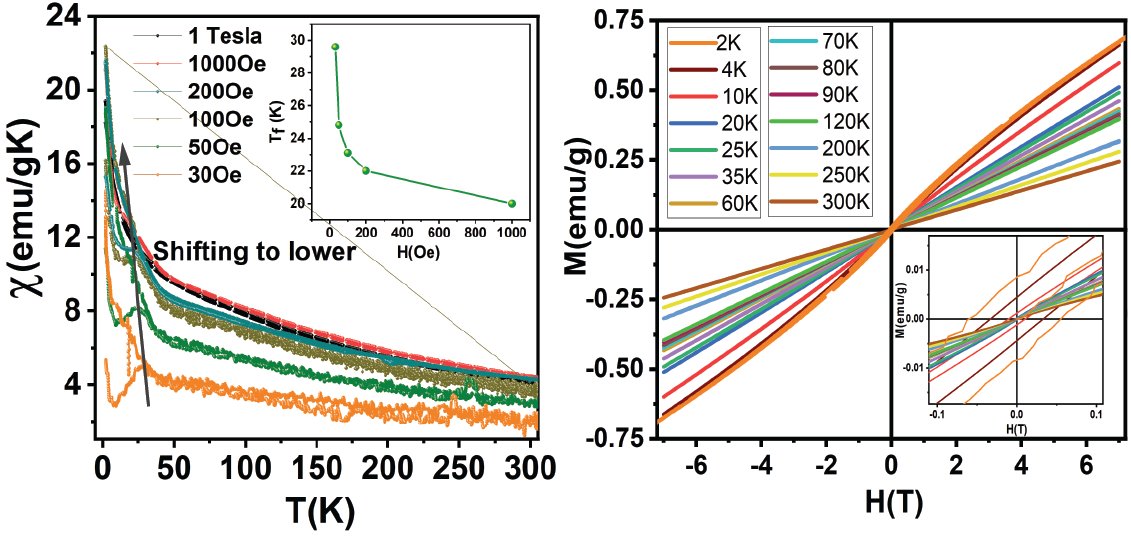}
	\caption{DC susceptibility measurement (a) ZFC and FC susceptibilities as a function of 
	temperature measured at different applied magnetic field, inset shows variation of T$_f$ with 
	applied magnetic field (b)isothermal magnetisation at various temperatures, inset shows hysterisis loop
	at lower temperatures below 20 K.}
    \label{Magnetisation}
\end{figure}

The low field DC $\chi$ from $2$ to $300$~K measured in zero-field-cooled (ZFC) and field-cooled (FC) modes at different applied magnetic 
fields $H$ is shown in Figure.~\ref{Magnetisation}(a).  In low fields $H = 30$~Oe we observe a cusp 
in ZFC at $T_{f} = 29$~K and a separation of the ZFC and FC data below this temperature \cite{gyani,young}. 
The $\chi$(T) measured at 
different magnetic fields is shown in Figure.~\ref{Magnetisation}(a).  The T$_f$ shifts towards lower values with magnetic field as shown in the inset of Figure.~ \ref{Magnetisation}(a) and the bifurcation 
between ZFC and FC also diminishes with field. 
These observations are consistent with a frozen spin glass (SG) state below T$_f$ \cite{Mydosh}.

Magnetic isotherm M(H) measurements were performed at various temperatures,
as shown in Figure\ref{Magnetisation}(b). At high temperatures, M(H) exhibits a nearly linear behavior,
consistent with the expected response in the paramagnetic (PM) 
region. However, as the temperature decreases, the magnetization curve develops
a noticeable curvature, which becomes more pronounced at lower temperatures, especially below 20 K.
At the lowest measured temperature of 2 K, a clear hysteresis is observed, with a coercive field
of approximately 1150 Oe shown in inset Figure\ref{Magnetisation}(b). Notably, even at an applied field of 9T, the magnetization does not reach saturation
unlike a ferromagnet, where spins align easily in one direction with applied field. 
The observation of hysterisis with no saturation till 9 T again suggests 
SG state, the spins get trapped in metastable energy states \cite{Mydosh,young}.
The seemingly hard magnetic properties of spin glasses
is usually explained by random anisotropy \cite{Mydosh}.

\begin{figure}[h]
    \centering
    \includegraphics[width=0.45\textwidth]{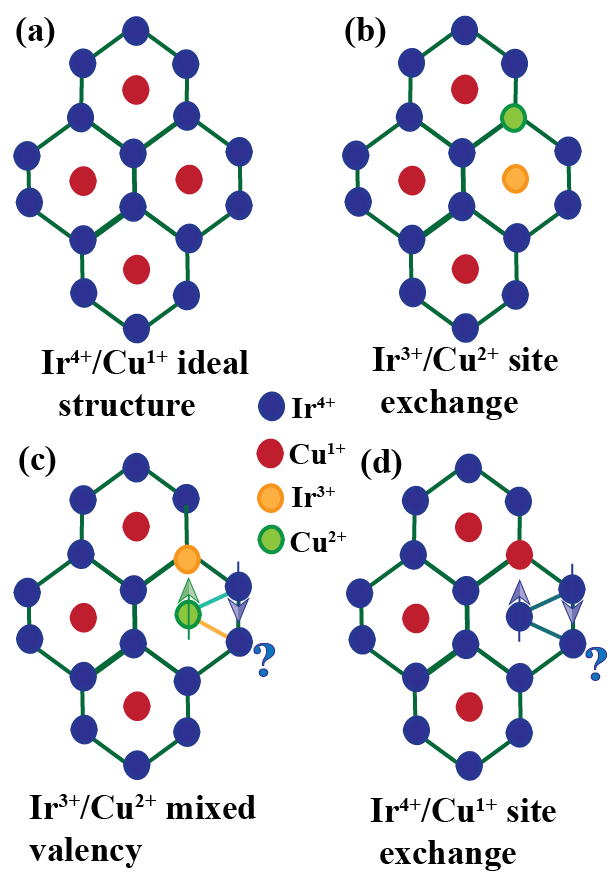}
     \caption{Schematic representation of possible types of chemical 
	disorders in the honeycomb plane. Cation disorder between 
	Cu\textsuperscript{1+}(red) and Ir\textsuperscript{4+}(blue), along with mixed
	valency involving Cu\textsuperscript{1+} (Ir\textsuperscript{4+}) and
	Cu\textsuperscript{2+}(Ir\textsuperscript{3+})(green(orange)), not only 
	introduces nonmagnetic impurities but also generates new triangular motifs.}
    \label{structure}
\end{figure}

To make sense of the observed frozen magnetic state we look to the structural disorder.  The magnetic site disorder and the ability of the octahedral coordination 
in the honeycomb layer to accommodate both Cu$^{1+}$ and Cu$^{2+}$ can lead to the formation of magnetic triangular motifs as shown in Fig.~\ref{structure}. 
These motifs can introduce frustrated magnetic interactions. Additionally, the competition between different exchange pathways, such as Ir$^{4+}$-O-Ir$^{4+}$ and Ir$^{4+}$-O-Cu$^{2+}$, likely plays a significant role in the emergence of glassy magnetic behavior at low temperatures.
The ordered magnetic moment per formula unit calculated from M(H) is quite small, only 0.54 $\mu$$_B$ at 9 T in 2 K isotherm M(H).

\begin{figure}[h]
    \centering
    \includegraphics[width=0.49\textwidth]{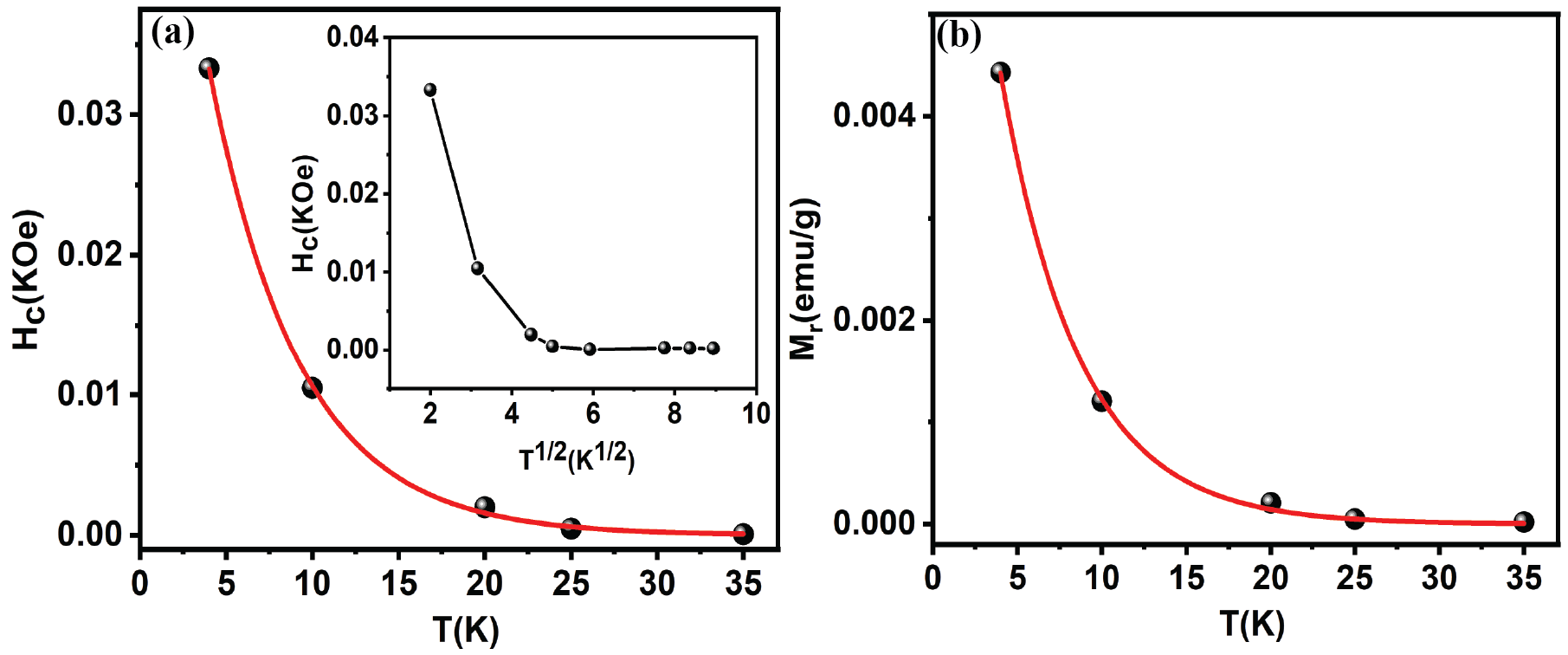}
	\caption{To differentiate between spin glass (SG), cluster glass, or superparamagnetic (SPM) behavior, the coercivity and retentivity vs. temperature plot follows:  (a) Empirical equation (7), with the inset demonstrating a non-linear behavior described by equation (6), ruling out SPM behavior.(b) Empirical equation (7), indicating that \CIO exhibits characteristics of either SG or cluster glass.}
	\label{MrHc}
\end{figure}

We would like to get a better understanding of the frozen magnetic state that we observe.  Either of spin-glass (SG), cluster spin-glass (CSG), or superparamagnetic (SPM) states can likely lead to the observed glassy behavior at low temperatures.
One of the ways to distinguish between these various states is through the temperature dependence of coercivity ($H_c$) \cite{Pandey}. 
The variations of $H_c$ and remanent magnetization ($M_r$) with temperature, determined from the $M(H)$ hysteresis loops, are shown in Fig.~\ref{MrHc}.

For non-interacting SPM systems, $H_c$ is expected to follow the temperature dependence:

\begin{equation}
H_c = H_{c0} \left( 1 - \frac{T}{T_B} \right)^{1/2}
\end{equation}

where $H_{c0}$ represents the coercivity at $T \to 0$, and $T_B$ is the blocking temperature,
defined as the temperature corresponding to the peak in the ZFC $\chi(T)$ curve.
As shown in the inset of Fig.~\ref{MrHc}, the $H_c$ vs $T^{1/2}$ plot exhibits a non-linear behavior,
ruling out the SPM blocking as the mechanism for the observed ZFC $\chi(T)$ peak.
Instead, the coercivity decreases exponentially with temperature as shown in the Fig.~\ref{MrHc}, following an empirical relationship commonly reported for spin-glass and cluster spin-glass systems below the freezing temperature $T_f$:

\begin{equation}
H_c (T) = H_c (0) e^{-\alpha T}
\label{eq:Hc}
\end{equation}

\begin{equation}
M_r (T) = M_r (0) e^{-\beta T}
\label{eq:Mr}
\end{equation}

where $H_c(0)$ and $M_r(0)$ represent the coercivity and remanent 
magnetization at $T = 0$ K, while $\alpha$ and $\beta$ are phenomenological parameters. 
The solid curves through the data in Fig.~\ref{MrHc} represent fits to the above Eqns.~\ref{eq:Hc} and \ref{eq:Mr}, yielding
$H_c(0) = 0.07$ KOe and $\alpha \approx 0.2$ K$^{-1}$ and 
$M_r(0) = 0.01$ KOe and $\beta \approx 0.2$~K$^{-1}$.

The observed exponential decay of $H_c$ and $M_r$ is consistent with either a spin-glass or a cluster spin-glass state \cite{Pandey}.

\section{Non equilibrium dynamics}
We now turn our attention to the non-equilibrium dynamic properties observed in the glassy state below $29$~K\@. 
We examine the key signatures such as the long-time relaxation of thermoremanent magnetization,
as well as memory and rejuvenation effects which are defining traits of spin glass behaviour \cite{Pandey,claudia,Pbag}.

\subsection{Magnetic relaxation measurements}
\begin{figure}[h]
    \centering
    \includegraphics[width=0.49\textwidth]{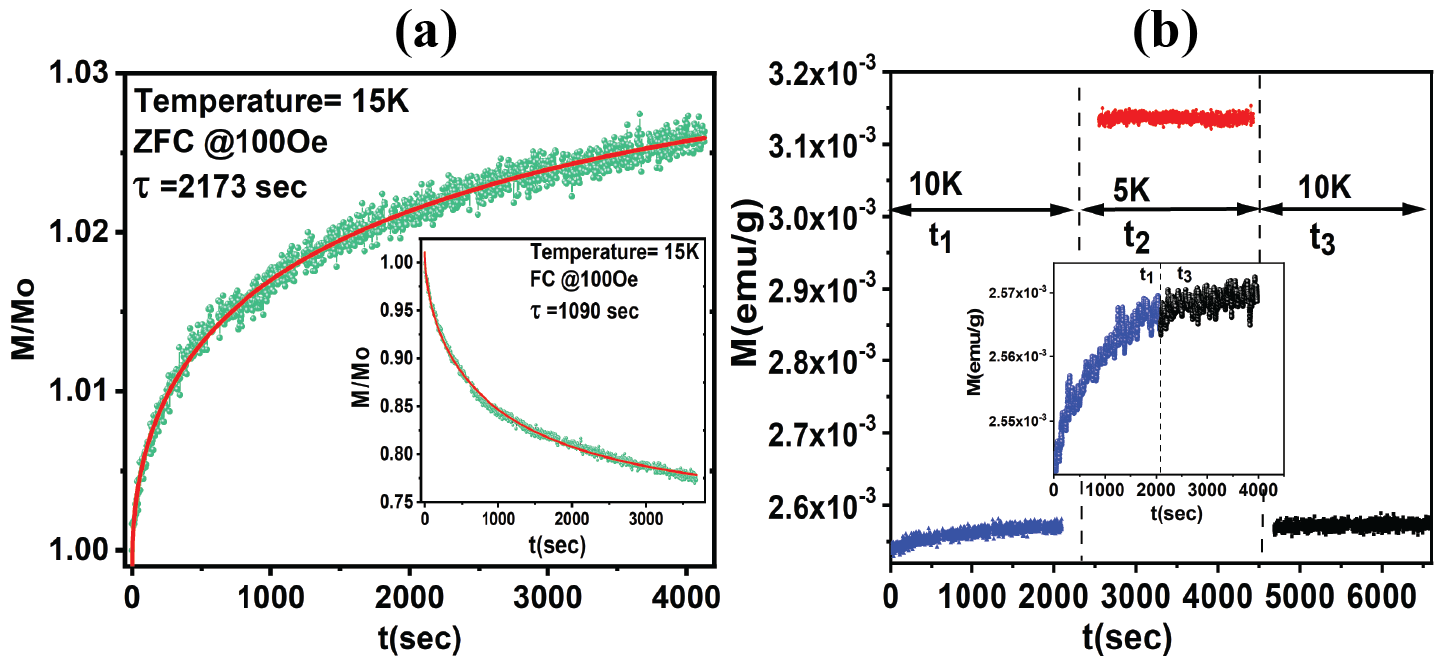}
	\caption{(a)Relaxation of the ZFC magnetization
	measured at T=15 K, inset shows FC relaxation for same temperature, 
	both fitted using strecthed exponential function(red curve)(b) Magnetic relaxation 
	measurements at 10 K with intermediate cooling to 5 K in an applied field of H = 500 Oe for ZFC,
	inset shows magnetization data from intervals 1 and 3 can be merged and seen to be continuous.}
\label{Relaxation}
\end{figure}

To investigate the relaxation behavior, time-dependent magnetization 
measurements were conducted at various temperatures below freezing temperature T$_f = 29$~K\@. 
To investigate the aging (ZFC relaxation) the sample was cooled in zero-field from above $T_f$ to $15$~K, and after waiting for a time $t_w = 1000$~s a magnetic field
of $100$~Oe was applied and the moment versus time was recorded.
The resulting time evolution of magnetization, $M(t)$, is shown in Fig.~\ref{Relaxation}.
 A gradual increase in magnetization over time, known as the magnetic aftereffect,
is observed. We can clearly see that the
magnetization does not saturate even after 1 hour. This is because,
in the glassy state, the moments are randomly frozen and it
takes a long time for the field to turn those spins along the
field direction. The $M(t,H)$ can be described by the standard stretched
exponential expression:

\begin{equation}
M(t, H) = M_0(H) + [M_{\infty}(H) - M_0(H)] \left[ 1 - \exp \left( -\frac{t}{\tau} \right)^{\alpha} \right]
\end{equation}

where $\tau$ represents the characteristic relaxation time, 
$\alpha$ is the stretching parameter (ranging between 0 and 1),
and $M_0$ and $M_{\infty}$ are the 
magnetization values at $t \approx 0$ and $t \rightarrow \infty$, respectively \cite{DNA,claudia,Chakrabarty}.
The best fit parameters obtained for the data at $15$~K are $\alpha = 0.48$ and $\tau = 2173$~s.
The values of  $\tau$ and $\alpha$ typically fall in the range for spin glasses reported
earlier \cite{Mydosh,SE,claudia,VF,Pbag,Pandey}.
The glassy behavior in the \CIO sample is further
supported by FC thermal remannent magnetization (TRM), as shown in Fig.~\ref{Relaxation} inset. For this measurement the sample was cooled from $300$~K to $15$~K in $100$~Oe and after waiting for $1000$~s the field was turned off and the moment
versus time was recorded.
The time evolution of magnetization can be satisfactorily modeled by Eq. (9). The solid red curves through the data in the Figure represent the best fit, and the obtained fit parameters are $\alpha = 0.58$; $\tau = 1090$~s at $15$~K\@.
These values are within the typical range reported for spin-glass systems and are similar to values obtained above from the ZFC relaxation analysis\cite{Mydosh,SE,claudia,VF,Pbag,Pandey,Sglass}.
\subsection{Memory effect}
The memory effect is another hallmark feature of spin glasses, reflecting their non-equilibrium 
dynamics and aging phenomena\cite{Mydosh,claudia,VF,Pbag,Pandey}. To observe this effect, the sample was
ZFC from the paramagnetic state down to 10 K, where a magnetic field of 500 Oe was applied. Similar to 
relaxation experiments, the magnetization was recorded as a function of time (interval 1). 
After 40 minutes, the temperature was rapidly decreased to 5 K and 
maintained for an additional 40 minutes (interval 2), before being increased back to 10 K (interval 3).

The $M$ versus $t$ for the three intervals is shown in Fig.~\ref{Relaxation}(b).  The inset shows that the magnetization data from interval-1 
and interval-3, both of which are measured at $10$~K seamlessly overlap, indicating that the system retains information 
about its prior state despite the intermediate cooling to $5$~K in interval-2.
The observed relaxation behavior can be understood within the framework of the
hierarchical model of the spin glass state \cite{HG,HG1}. According to this model, 
at a given temperature \( T_0 \), the free-energy landscape consists of 
multiple interconnected valleys. When the sample is cooled from \( T_0 \) to \( T_0 - \Delta T \), each
existing valley further subdivides into smaller sub-valleys, following the
hierarchical organization. If \( \Delta T \) is large, the energy barriers 
between different valleys become significantly high, 
restricting relaxation to transitions within the newly formed sub-valleys. Upon reheating the 
system back to \( T_0 \), these sub-valleys merge,
restoring the original energy landscape and thereby demonstrating the memory effect 
in the decreasing temperature cycle.

\subsection{AC Magnetisation}
AC susceptibility is a tool for investigating the dynamics of spin glasses.
A sharp 
cusp in the real part of the AC susceptibility $\chi'$ is usually observed at the freezing temperature ($T_f$) \cite{AC1970,AC19702,Mydosh}.  The frequency dependence can reveal important information about the glassy state.  
Below $T_f$, $\chi'$ decreases with increasing
frequency, whereas in the paramagnetic regime above $T_f$, all
frequency-dependent curves overlap. A key characteristic of spin glasses
is the upward shift of $T_f$ with increasing frequency. In contrast,
systems with long-range ferromagnetic or antiferromagnetic order
exhibit such frequency-dependent shifts only at much 
higher frequencies, typically in the MHz range \cite{Mydosh}.

\begin{figure}[h]
    \centering
    \includegraphics[width=0.55\textwidth]{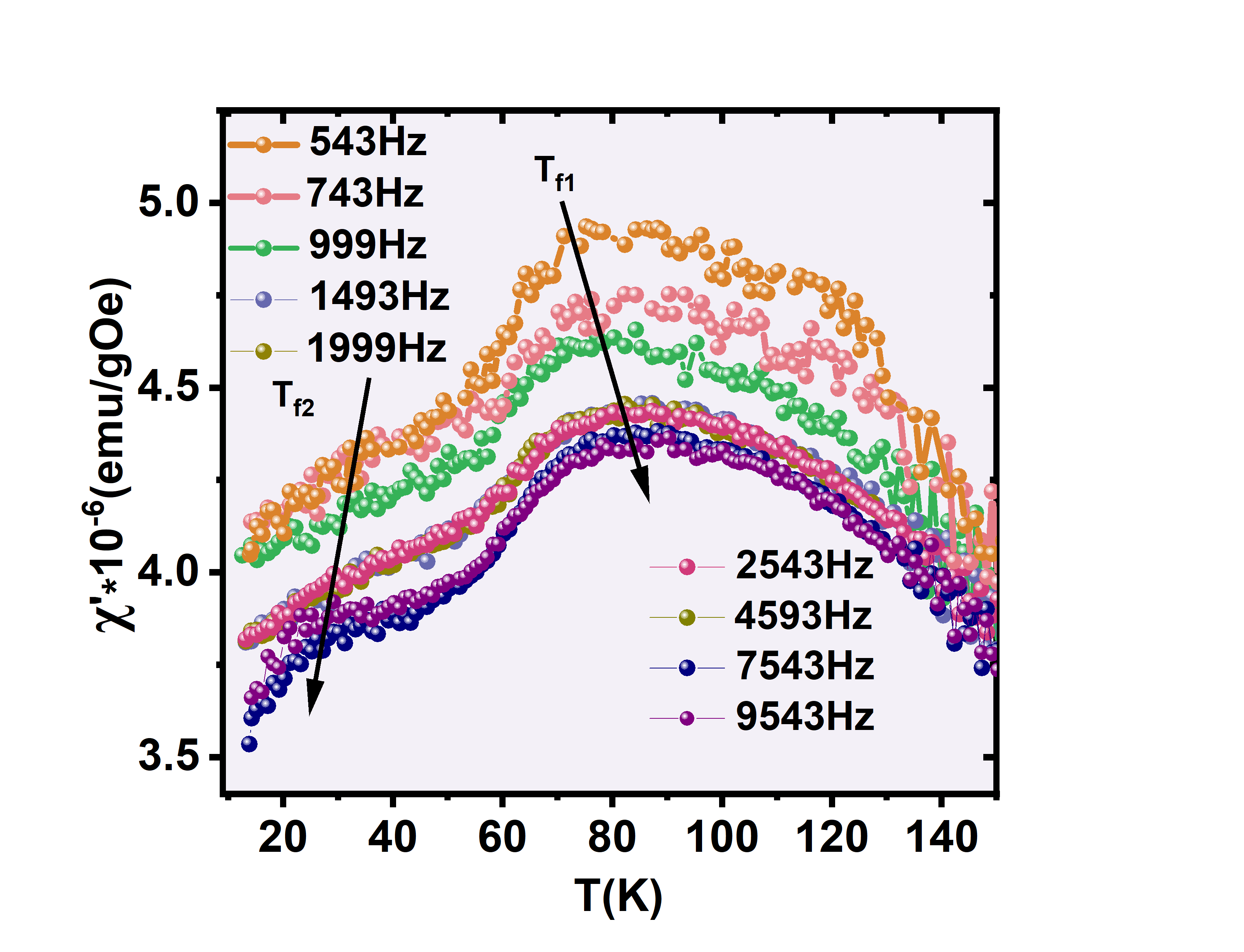} 
    \caption{Temperature dependence of the real part of AC susceptibility ($\chi'$) measured at different frequencies. A primary frequency dependent peak is observed at $T_{f1}$ with a smaller shoulder at $T_{f2}$.}
    \label{ACchi}
\end{figure}

To investigate the spin glass (SG) dynamics in Cu$_2$IrO$_3$, AC susceptibility measurements
were performed in the frequency range of 543 Hz to 9543 Hz, using a fixed excitation
field of $H_{\text{ac}} = 10$ Oe, after cooling the sample in 
zero field. The temperature dependence of $\chi'$ is shown in Fig.~\ref{ACchi}, where a primary peak at $T_{f1}$ is observed along with 
a smaller shoulder at a lower temperature $T_{f2}$.
We note that in DC $\chi$ an anomaly was only observed at $T_{f2}$.  
Both peaks observed in ac $\chi$ are quite broad unlike the sharp cusps normally observed in canonical spin-glasses. 
Such broadened peaks suggest the freezing of correlated spin clusters rather than isolated spins.
The interactions within these clusters create a distribution of relaxation times,
leading to a smoother transition instead of a sharp freezing point \cite{Mydosh}.
The peak at $T_{f1}$ shifts to higher temperatures with increasing frequency, whereas the shoulder peak at $T_{f2}$ responds only at higher frequencies as shown by the arrows in Fig.~\ref{ACchi}.

The peak behavior of \( T_{f1} \) can be analyzed using
the Mydosh parameter, which quantifies the shift in freezing temperature per decade of frequency and is defined as:  

\[
\delta T_f = \frac{\Delta T_f}{T_f \cdot \Delta \log_{10} \nu}
\]

where \( \Delta T_f = T_f(\nu_1) - T_f(\nu_2) \) and
\( \Delta \log_{10} \nu = \log_{10} (\nu_1) - \log_{10} (\nu_2) \). 

We obtain \( \delta T_f \approx 0.089 \) which falls between the range 
characteristic of canonical spin glasses (\( \delta T_f \sim 0.005 \) for CuMn, \( \delta T_f \sim 0.0045 \) AuMn) 
and non-interacting ideal superparamagnets (\( \delta T_f \sim 0.28 \) for
holmium borate glass \( \alpha \)-Ho\(_2\)O\(_3\)(B\(_2\)O\(_3\))) \cite{AuMn,Mydosh}. 
The obtained \( \delta T_f \) is notably larger than that of metallic spin glasses but is comparable to the value found in insulating spin glasses \cite{Mydosh}. 
The much larger value of \( \delta T_f \) compared to canonical spin glasses and the broad anomalies at $T_f$ in ac $\chi$ suggest that the system may be best described as a cluster glass, where correlated spin clusters govern the magnetic dynamics.  

To distinguish between the spin-glass, cluster glass,
and superparamagnetic states, three empirical laws are commonly used to analyze the frequency dependence of $T_{f}$:  

1. Arrhenius Law:  
   \begin{equation}
   \tau = \tau_0 \exp \left(\frac{E_a}{k_B T_f}\right)
   \end{equation}
   This model is typically applicable to non-interacting superparamagnets or weakly interacting magnetic entities.  

2. Vogel-Fulcher (VF) Law:  
   \begin{equation}
   \tau = \tau_0 \exp \left[\frac{E_a}{k_B(T_f - T_{g})}\right]
   \end{equation}
   This approach accounts for interactions between spins.  

3. Critical Scaling (CS) Approach:  
   \begin{equation}
   \tau = \tau_0 \left[ \frac{T_f}{T_{g}} - 1 \right]^{-z\nu}
   \end{equation}
   
Here, \( \tau \) represents the relaxation time (inverse of the AC frequency)
which describes the dynamical fluctuation time scale. \( T_f \) is the experimentally 
determined freezing temperature, \( \tau_0 \) is 
the characteristic spin relaxation time
(reflecting the timescale of a single spin flip),
and \( E_a \) represents the activation energy barrier
separating metastable states, and \( T_{g} \) 
is the static freezing temperature as \( \nu \) approaches
zero, while the exponent \( z\nu \) characterizes the divergence of the correlation length, defined as  
\begin{equation}
\zeta = \left(\frac{T_f}{T_{g}} - 1\right)^{-\nu}
\end{equation}
and \( k_B \) is the Boltzmann constant.  

\begin{figure}[h]                                            
\centering                                     \includegraphics[width=0.49\textwidth]{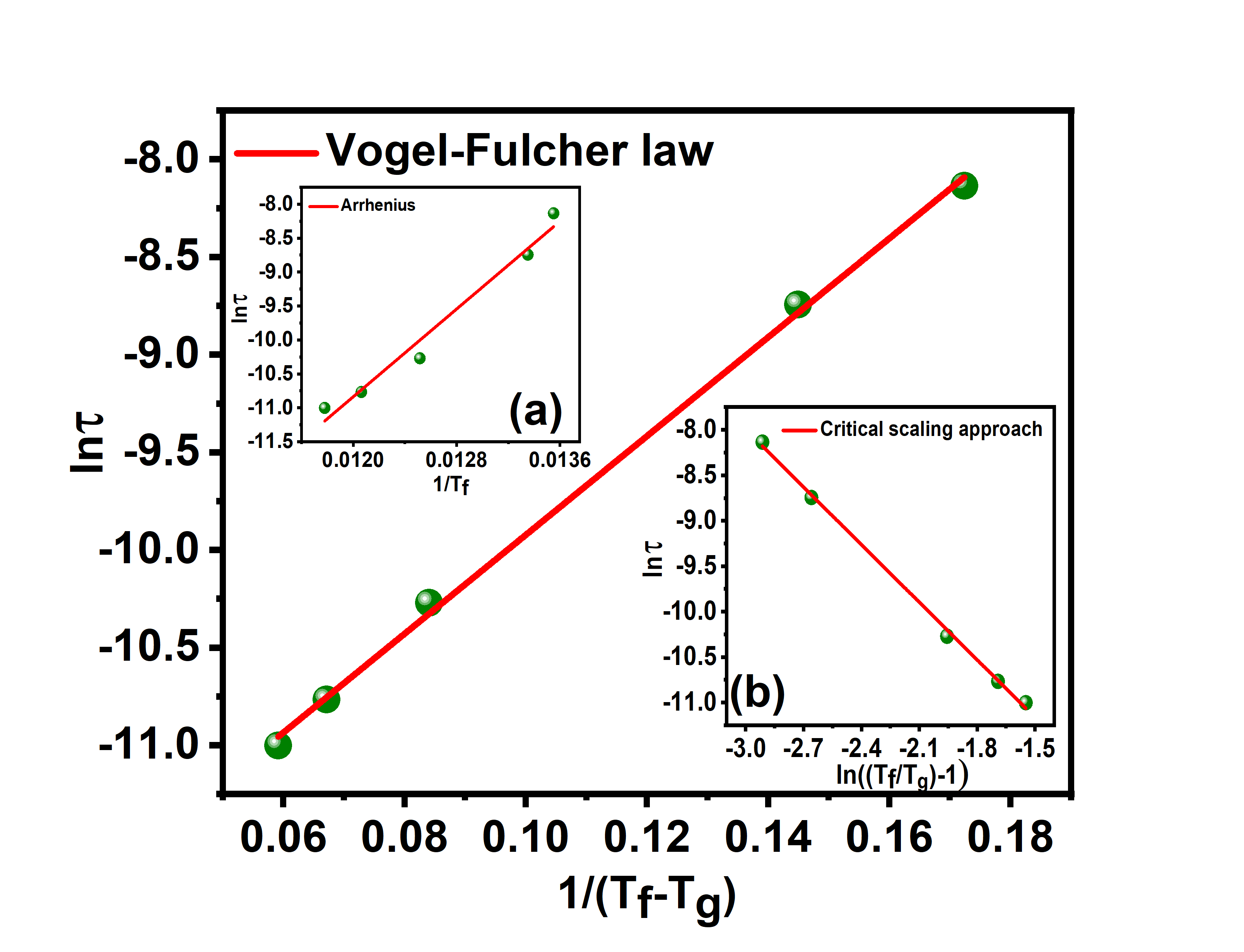}      
\caption{Fitting the frequency dependence of T$_f$ showing 
	the Vogel-Fulcher law and the critical scaling approach (inset below) both fit
	equally well, but the Arrhenius law fails (inset above).}
    \label{vogelfulcher}
\end{figure}

The results of parameters obtained from fits to each of these models is summarized in Table~\ref{VFparameters}.
The inset of Figure \ref{vogelfulcher} (a) demonstrates 
that the Arrhenius law fails to describe the frequency shift in \CIO. 
It does not fit across the entire frequency range, 
and the extracted parameters (\( \tau_0 \approx 8.28 \times 10^{-14} \) s, \( E_a/k_B \approx 1615\)~K) 
seems unphysical \cite{Pbag}.
We use this to rule out a superparamagnetic state, which also exhibit frequency-dependent \( T_f \) behavior
but follow the Arrhenius law. The inability of the Arrhenius model
to capture the dynamics suggests that the observed behavior 
is not due to independent spin flips but rather arises from cooperative interactions among spin clusters.  

In contrast, both the Vogel-Fulcher and critical-scaling laws, which incorporate the interaction of spins, provide a reasonable
fit over the entire frequency range (Figure \ref{vogelfulcher} main panel and inset (b)), supporting the presence of cooperative spins. 
The Vogel-Fulcher plot yields the parameters \( E_a/k_B \) and \( \tau_0 \) obtained 
from the slope and intercept of the linear fit. The obtained values are summarised in table \ref{VFparameters}. 

Similarly, a plot of \( \ln(\tau) \) versus \( \ln(T_f / T_g - 1) \), with \( T_g = 70 \) K (determined from the best fit of data) for critical-scaling 
approach (inset Figure \ref{vogelfulcher}(b)),
yields parameters, \( \tau_0 \) and \( z\nu \) from intercept and slope of linear fit. 
In conventional SG systems, \( z\nu \) typically falls within the range $4$ to $12$, while \( \tau_0 \) ranges from \( 10^{-10} \) to \( 10^{-13} \) s for canonical SG and from \( 10^{-6} \) to \( 10^{-10} \) s for cluster SG \cite{claudia,Pbag,twopeak}.  

The obtained spin relaxation time for \CIO, \( \tau_0 \approx 10^{-7} \) s and  \( \tau_0 \approx 10^{-6} \) s 
from critical scaling and Vogel Fulcher model respectively, 
falls within the characteristic range for cluster glasses suggesting that in \CIO
spin dynamics occurs in a slow manner, due to the presence
of interacting clusters rather than individual spins. The values obtained for $\tau_0$ are similar to those reported in many other cluster-
SG systems such as Fe$_2$O$_3$, K$_3$CrO$_3$ and Ni doped La$_{1.85}$Sr$_{0.15}$CuO$_4$ \cite{Fe2O3,K3CrO3,todiff}.
The extracted critical exponent \( z\nu \approx 2.1 \) value is also
less than the range for usual spin glasses. \cite{Chakrabarty,K3CrO3}  
All these observations taken together suggest that the low temperature frozen state is a cluster-glass instead of a spin-glass.
 The ratio 
\( k_BT_g / E_a \) obtained from the fit to the Vogel-Fulcher model serves as an indicator of interaction
strength between dynamic entities with low values ($\leq 1$) suggesting weak coupling and higher
values ($> 1$) implying strong coupling \cite{vgf}. In \CIO, 
\( k_BT_g/E_a \approx 2.72 \), placing the system in
the moderate to strong interaction regime. This suggests moderate interactions 
among the magnetic clusters.

 \begin{table}[h]
    \centering
    \renewcommand{\arraystretch}{1.1}
    \setlength{\tabcolsep}{2pt}
    \small
    \begin{tabular}{lccc}
        \hline\hline
        \textbf{Parameter} & \textbf{Arrhenius} & \textbf{Vogel-Fulcher} & \textbf{Critical Scaling} \\
        \hline
        $\tau_0$ (s) & $8.3\!\times\!10^{-14}$ & $3.9\!\times\!10^{-6}$ & $6.0\!\times\!10^{-7}$ \\
                     & $\pm 3.9\!\times\!10^{-14}$ & $\pm 0.25\!\times\!10^{-6}$ & $\pm 0.69\!\times\!10^{-7}$ \\
        $E_a / k_B$ (K) & $1615 \pm 147$ & $25.3 \pm 0.5$ & -- \\
        $T_{SG}$ (K) & -- & $68$ & $70$ \\
        $z\nu$ & -- & -- & $2.10 \pm 0.05$ \\
        \hline\hline
    \end{tabular}
    \caption{Comparison of parameters obtained from Arrhenius, Vogel-Fulcher, and Critical Scaling models.}
    \label{VFparameters}
\end{table}

Hence our DC and AC susceptibility results conclude that at temperature \( T_{f1} \),
dynamically fluctuating AFM clusters are formed \cite{dynamic,dynamic2,VF,dynamic4}.
The correlated spins within these clusters retain sufficient energy at such high temperatures, 
preventing static freezing.
Dynamic scaling using VF and CS law indicates the correlated behaviour of spins and the 
strong inter-cluster interactions
hinder independent spin freezing, resulting in the cooperative slowing down of relaxation time.
As the temperature decreases below \( T_{f1} \), clusters start influencing each
other more strongly and become harder to reorient independently eventually driving the system towards cluster freezing.
The onset of strictly frozen clusters is observed below 29 K in 
dc susceptibility, memory and relaxation measurements. 
Frequency response of \( T_{f2} \) shoulder at higher frequency only in ac susceptibility 
also confirmed the emergence of static frozen state. 
Clusters with varying relaxation times respond differently to different frequencies,
the response at higher frequencies indicates the presence of 
clusters with time scales comparable to the applied frequency.
Energetically at T$_{f2}$
the cluster behaves as a single large unit, the effective energy barrier is higher enough to
flip the correlated spins and clusters freeze together.
Since DC susceptibility measures slow,
static magnetization changes, it signals the onset of a non-ergodic state at 29K but
does not indicate any transition for dynamically fluctuating clusters around 80K.
Furthermore, below 20K, a fully frozen state is observed, 
as evidenced by the emergence of hysteresis, which confirms
irreversibility indicating that the clusters remain frozen even when the applied field is cycled.

\section{Summary}
We have synthesized Cu$_2$IrO$_3$ by an ion exchange reaction of 48 hours at a temperature of 320$\degree$C. 
Using XRD, EXAFS and PDF we quantify approximately 25\% atomic disorder within the honeycomb layers of Cu$_2$IrO$_3$. XANES and XPS revealed mixed valence states of Cu and Ir following $\text{Cu}^{1+} + \text{Ir}^{4+} \rightarrow \text{Cu}^{2+} + \text{Ir}^{3+}$.
This site disorder and mixed valence introduce complex magnetic interactions in \CIO,
characterized by competing antiferromagnetic (AFM) interactions and 
frustrated triangular motifs. The AC and DC susceptibility, memory, and relaxation studies indicate the formation of dynamically fluctuating, strongly correlated AFM clusters at approximately $80$ K with moderate inter-cluster interactions. Below temperature $29$ K, freezing of these clusters sets in. 
These findings highlight the crucial role of synthesis conditions in governing the 
structural and magnetic properties of complex oxides. Precisely
tuning disorder may be a pathway to engineering emergent quantum phases in strongly correlated oxides.
\section{Acknowledgement}
We thank the X-ray, SEM, and the liquid Helium facilities at IISER Mohali.  Y. S. acknowledges support from SERB project CRG/2022/000015 and STARS project STARS-1/240.
We acknowledge DESY (Hamburg, Germany), a member of the Helmholtz Association HGF, for the provision of experimental facilities. Parts of this research were carried out at PETRA III, beamline P64, P02.1 and P04 and we thank Dr. Aleksandr Kalinko for beamline guidance at P64 beamline. PY would like to thank Dr. Supriyo Majumdar, Dept. of Material Science and Engineering, Northwestern University, Evanston, Illinois for his useful discussion in analyzing EXAFS section. We would like to thank Dr. R. J. Choudhury UGC-DAE Consortium for Scientific Research, Indore for X-ray photoemission spectroscopy measurements. Dr. Rajeev Rawat, Kranti Kumar and Somya Shephalika, UGC-DAE Consortium for Scientific Research, Indore, are acknowledged for isothermal MH measurements. We thank Dr. Ashna Bajpai (IISER Pune) for fruitful discussions related to the magnetic study.

\bibliographystyle{unsrt}
\bibliography{CIO_CG1.bib}

\end{document}